\documentclass{emulateapj}
\shorttitle{Spin-down of 4U 2206+54}
\shortauthors{Finger, Ikhsanov, Wilson-Hodge \& Patel}
\usepackage{graphicx}

\newcommand{\be}{\begin{equation}}
\newcommand{\ee}{\end{equation}}
\newcommand{\bdm}{\begin{displaymath}}
\newcommand{\edm}{\end{displaymath}}

\begin{document}

\title{Spin-Down of the Long-Period Accreting Pulsar 4U 2206+54}
\author{Mark H. Finger\altaffilmark{1,2}, Nazar R. Ikhsanov\altaffilmark{1,3,4}
        Colleen A. Wilson-Hodge\altaffilmark{3}, and Sandeep K. Patel\altaffilmark{5}}
\altaffiltext{1}{National Space Science and Technology Center, 320 Sparkman Drive, Huntsville, AL} 
\altaffiltext{2}{Universities Space Research Association, 6767 Old Madison Pike, Suite 450, Huntsville, Alabama 35806}
\altaffiltext{3}{Space Science Office, VP 62, NASA Marshall Space Flight Center, Huntsville, AL 35812} 
\altaffiltext{4}{Pulkovo Observatory, 196140 St.\,Petersburg, Russia}
\altaffiltext{5}{Optical Sciences Corporation, 6767 Old Madison Pike, Suite 650, Huntsville, Alabama 35806}

\begin{abstract}
4U 2206+54 is a high mass X-ray binary which has been suspected to contain a neutron
star accreting from the wind of its companion BD +53$\degr$ 2790. \citet{Reig-etal-2009}
have recently detected 5560 s period pulsations in both RXTE and INTEGRAL observations
which they conclude are due to the spin of the neutron star. We present observations
made with Suzaku which are contemporaneous with their RXTE observation of this source.
We find strong pulsations at a period of $5554 \pm 9$~s in agreement with their results.
We also present a reanalysis of BeppoSAX observations of 4U 2206+54 made in 1998, in
which we find strong pulsations at a period of $5420 \pm 28$ seconds, revealing a
spin-down trend in this long-period accreting pulsar. Analysis of these data suggests
that the neutron star in this system is an accretion-powered magnetar.
\end{abstract}


\section{Introduction}
First observed by Uhuru and Ariel V, the low luminosity
($\sim$10$^{35}$erg cm$^{-2}$s$^{-1}$)
hard x-ray source 4U 2206+54 is one of a handful of well studied galactic
high mass x-ray binaries for
which the nature of the accreting compact star is uncertain.

Optical and UV spectroscopy have showed the optical counterpart, BD +53$\degr$ 2790  to
be an unusual O9 active star with strong wind resonance lines in the ultraviolet
\citep{Negueruela01}.  Several authors have concluded from the high column depth and the
flaring observed in the x-rays that the compact object is accreting from a strong
stellar wind. Rib\'{o} et al. (2006) determine from the an IUE spectra a relatively low
wind velocity of $\sim$350 km s$^{-1}$.

Analysis of the first five years of the RXTE ASM light-curve of 4U 2206+54 revealed a
periodic modulation at a period of 9.57 days, which was interpreted as the orbital
period of the binary system \citep{Corbet01}. Radial velocity measurements were made for
BD +53$\degr$ 2790, however it was not possible to independently determine the orbital
period from them \citep{Blay05b}. Recent analyses of the Swift/BAT and RXTE/ASM
light-curves have thrown this picture into confusion \citep{Corbet07}. The recent data
shows a modulation with a period of 19.25 days -- twice what was previously thought to
be the period. If 19.25 days is the orbital period, then the profile of the orbital
modulation has evolved from having two peaks per orbit to one peak per orbit.

Using observations extended over 7 days from the Proportional Counter Array (PCA)on the
Rossi X-ray Timing Explorer (RXTE) \citet{Reig-etal-2009} discovered pulsations at a
period of $5559\pm3$~s. Detecting pulsations in this period range from a low Earth
orbit is usually difficult because the pulse period is near the orbital period,
resulting in poor sampling of pulses. Their data however include several intervals
which were uninterrupted by Earth occultation. They also confirmed this
discovery using observations from IBIS on the International Gamma-ray Astrophysics Laboratory
(INTEGRAL) which has a high Earth orbit with a 3 day period.

In section 2 we present results from new observations of 4U 2206+54 we have made with
Suzaku. These observations were contemporaneous with those of RXTE
\citep{Reig-etal-2009}. We find strong pulsations at a pulse period of $5554\pm9$~s, in
agreement with the RXTE results. In section 3 we present a reanalysis of the BeppoSAX
observations of 4U 2206+54 which show strong pulsations at a period of $5420\pm28$~s, in
section 4 we discuss simulations of Suzaku and BeppoSAX data, in section 5 the EXOSAT 
observations of 4U 2206+54, and in section 6 we
briefly analyze the evolutionary status of the system and the nature of its primary
component. We conclude that the primary component of the system is a neutron star, which
has a huge ($\sim (3-5) \times 10^{15}$\,G) surface field and is accreting material onto
its surface from a disk. In this light 4U~2206+54 appears to be the first
accretion-powered magnetar identified so far.

\begin{figure*}[!t]
\centerline{\includegraphics[width=7.0in]{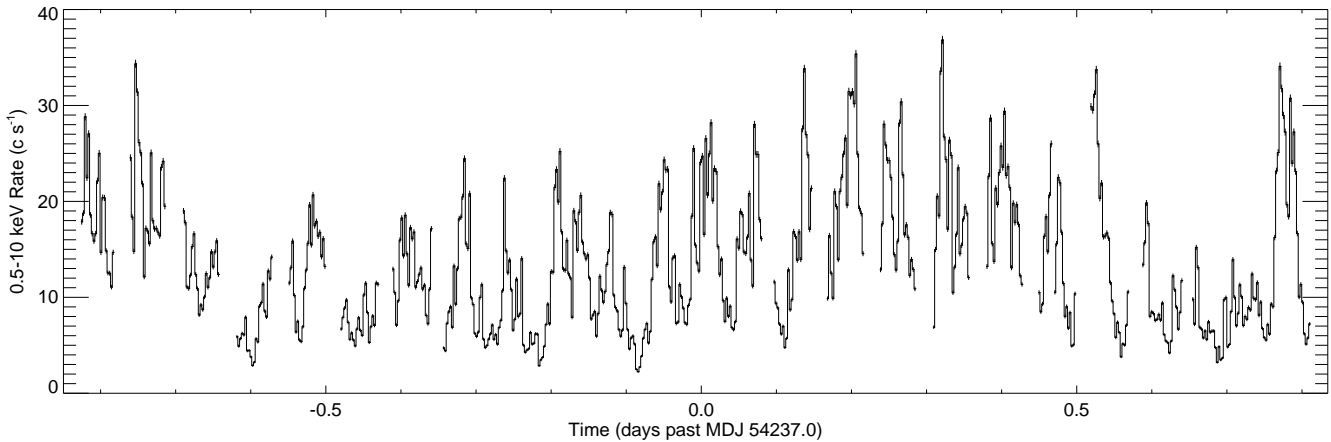}}
\caption{light-curve of 4U 2206+54 from Suzaku XIS (XIS0, XIS1 and XIS3 summed) in the 0.5 to 10 keV
energy band.\label{Suzaku_lc}}
\end{figure*}

\section{Analysis of the Suzaku light-curve}

We obtained an observation of 4U 2206+54 with Suzaku with the primary objective of
searching for long-period pulsations with the X-ray Imaging Spectrometer \citep[XIS;
][]{Komaya07}. To minimize the number and size of data gaps the observation was time
constrained to a phase of the space-craft precession cycle were the source was always
above the Earth horizon. The observation occurred 2007 May 16 to 17 (MJD
54236.176-54237.813), lasting for 141 ks, with a total exposure for the XIS of 114.8 ks
after filtering. A 20.9 ks duration portion of this observation (MJD
54236.861-54236.110) overlaps with the RXTE observations presented by
\citet{Reig-etal-2009}.

The XIS observations were taken in 1/4 window mode to avoid potential problems with
pile-up. Our analysis began with the version 2.0.6.13 processing data (which has aspect
correction for thermal wobbling).  The Charge Transfer Inefficiency (CTI) 
correction was remade and the event data
cleaned using the standard filtering with the exception that the lower bound on the day
side earth limb elevation was reduced from 20 degrees to 17.5 degrees (which did not
appear to increase the broad-band background). Events were selected from the active
telescopes (XIS0, XIS1, and XIS3) using 208\arcsec\, x 261\arcsec\, rectangular extraction
regions centered on the source, with the long axis aligned with the 1/4 window. As a limit
on the background, events were also extracted from each telescope in regions of the same
area as distant as possible from the source. The average rate summed over the telescopes
was 13.9 c s$^{-1}$ for the source regions and 0.125 c s$^{-1}$ for the background
regions.

The XIS light-curve in the 0.5-10.0 keV range with $\sim$\,200\,s binning is shown in Figure
\ref{Suzaku_lc}. In order to avoid a loss of data, unequal bins sizes were used. Each
good-time interval was divided into equal width bins, with the number of bins chosen to result
in the bin width closest to 200\,s. There appears to be pulsation with a period near 0.06
days, which is most evident in the spacing of the minima. Also evident is strong variablity at
much longer timescales.

\subsection{Pulse Search Statistic}

\begin{figure}[b!]
\centerline{\includegraphics[width=3.4in]{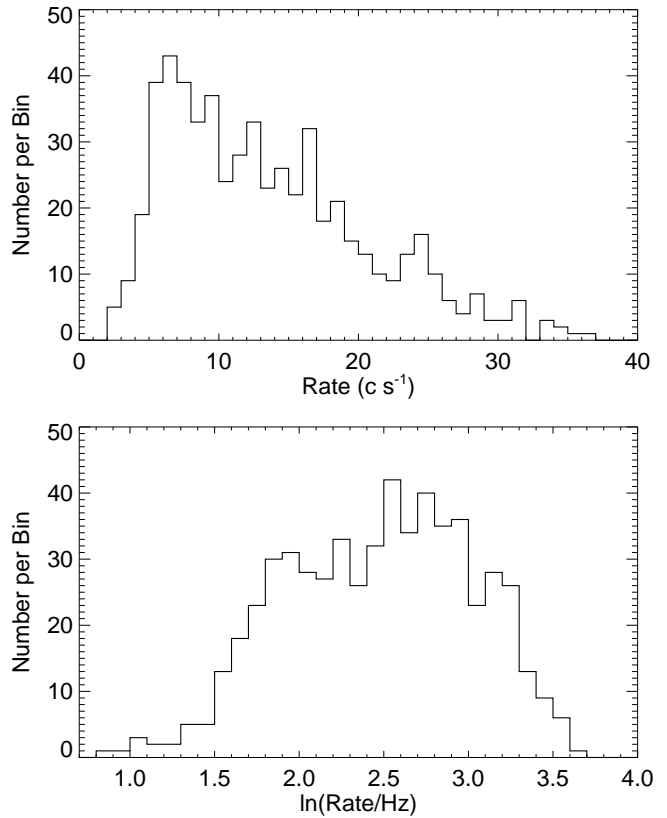}}
\caption{Distribution of the Suzaku XIS count rate (top panel),
and of its natural log (bottom panel).\label{Distributions}}
\end{figure}

To search for pulsation and determine its period we have developed a test statistic, which 
like the Lomb test statistic is based on hypothesis testing, but which can account for the
low frequency variablity and other characteristics of the observations. 

The Lomb method \citep{Lomb75}, a standard technique for detection of pulsations in
unevenly sampled data, assumes a null hypothesis of Gaussian white noise, and a alternate
hypothesis of Gaussian white noise plus a sinusiod.  
For 4U 2206+54 the assumption of Gaussian white noise is inappropriate. 
\citet{Reig-etal-2009} show that the power spectra of
the flux in the 2-20 keV band are power-laws above 1 mHz, with an index between 1.5 and 1.9.
The light curve in Figure \ref{Suzaku_lc} shows both short term flaring and strong
long period ($\sim$\,day) variability, suggesting this red noise spectra extends to lower frequencies.
In addition to the power-spectrum not being white, the distribution of the rates is far from being
Gaussian. In the top panel of Figure \ref{Distributions} we show the distribution of the rates measured
with Suzaku XIS, which is more exponentional than Gaussian. In the bottom panel we show the distribution
of the natural log of the rate, which has a more symmetric distribution which is closer to a Gaussian.

For a more realistic null hypothesis we will assume that the natural log of the flux is
Gaussian noise, with a power spectrum of the form
\begin{equation}
\mathcal{S}(f) = s_0 f_0^\Gamma(f^2+\alpha^2)^{-\Gamma/2} \label{powspec}
\end{equation}
where $s_0$, $\alpha$ and $\Gamma$ are adjustable constants and $f_0$ is a fixed reference frequency.
The covariance of the natural logs of rates $k$ and $l$ is then
\begin{equation}
V_{kl} = \sigma_k^2\delta_{kl}+\int_0^\infty \mathcal{S}(f) {\rm sinc}^2(\pi f \tau)
      \cos( 2\pi f [t_k-t_l]) df
\label{covar}
\end{equation}
where $\sigma_k$ is the error on $y_k = {\rm ln}(r_k \times 1{\rm~s})$, $\tau$ is the average
rate bin width, and the $ {\rm sinc}^2$ term accounts for the effect of binning. The likelihood
of the  null hypothesis is then given by
\begin{equation}
\mathcal{L}_0 = {\rm Det}(2\pi V)^{-\frac{1}{2}}
      \exp(-\case{1}{2}\sum_{kl} [y_k-\mu] V^{-1}_{kl} [y_l-\mu])
\label{nullhypo}
\end{equation}
where $\mu$ is the mean natural log rate. In practice it is simpler for us to use the Cash statistic
\citep{Cash79} $\mathcal{C} = -2{\rm ln}\mathcal{L}$, giving
\begin{equation}
   \mathcal{C}_0 = \sum_{kl} (y_k-\mu) V^{-1}_{kl} (y_l-\mu)-{\rm ln}({\rm Det}(V))+const.
\end{equation}
which is to be minimized with respect to $s_0$, $a_0$, $\Gamma$ and $\mu$.

For the alternative hypothesis the natural log of the flux is Gaussian noise with a power
spectra of the same form, plus a sinusoidal profile. The resulting Cash statistic is
\begin{equation}
   \mathcal{C}_1 = \sum_{kl} (y_k-p_k) V^{-1}_{kl} (y_l-p_l)-{\rm ln}({\rm Det}(V))+const.
\end{equation}
where the pulse profile $p$, is given by
\begin{equation}
   p_k = \mu + a\cos{2\pi f t_k}+ b\sin{2\pi f t_k} ~. \label{profile}
\end{equation}
$\mathcal{C}_1$ is to be minimized with respect to $s_0$, $a_0$, $\Gamma$, $\mu$, $a$,
and $b$ for a given trial frequency $f$. The test statistic analogous to the
$\Delta\chi^2$ is
\begin{equation}
   \Delta\mathcal{C} = \min(\mathcal{C}_0)-\min(\mathcal{C}_1)~.
\end{equation}
In the appendix we explain how the calculation of this statistic is implemented numerically. In Section 4 we
describe how results from this statistic compare with the Lomb method. 

\begin{figure}[!b]
\centerline{\includegraphics[width=3.4in]{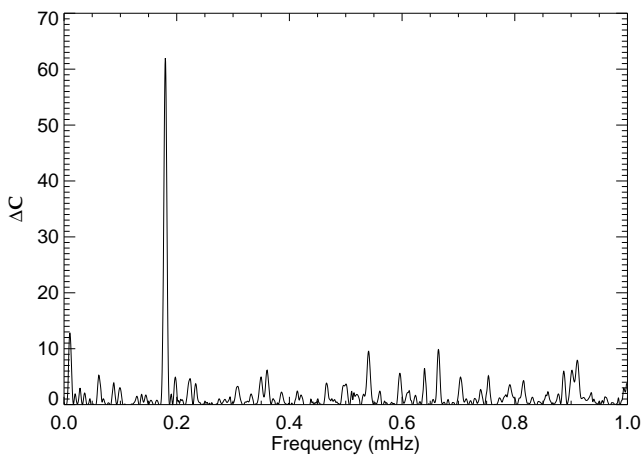}}
\caption{Delta Cash statistic versus frequency for the Suzaku XIS light curve.\label{Suzaku_dc}}
\end{figure}

\subsection{Suzaku Timing Results}

In Figure~\ref{Suzaku_dc} we show $\Delta\mathcal{C}$ for the Suzaku XIS light curve.
The false detection probability of the highest peak is $5\times 10^{-12}$ including the
number of search trials within the 0-1.0\,mHz interval. 

\begin{deluxetable*}{lcccccc}[!t]
\tablewidth{0pt}
\tablecolumns{7}
\tablecaption{Period and Power Spectra Parameter Estimates\label{param_tab}}
\tablehead{
\colhead{Observation} & \colhead{Mid-Time} & \colhead{Period} & \colhead{$S_0$\tablenotemark{a}} & \colhead{$\alpha$} & \colhead{$\Gamma$} & \colhead{Harmonics\tablenotemark{b}} \\
          \colhead{}  & \colhead{(MJD)}    & \colhead{(s)}    & \colhead{(Hz$^{-1}$)}            & \colhead{($10^{-5}$ Hz)} & \colhead{}   & \colhead{} }
\startdata
Suzaku XIS             &  54237.0       &   $5554\pm9$    &     $31.6 \pm 1.9$ &   $2.2\pm1.5$\tablenotemark{c}  &$1.10 \pm 0.08$ & 3  \\
BeppoSAX MECS          &  51141.1       &   $5420\pm28$   &     $21.4 \pm 3.6$ &   $< 12$\tablenotemark{d}       &$1.14 \pm 0.18$ & 2
\enddata
\tablenotetext{a}{Power spectra normalization at frequency $f_0 = 1$mHz.}
\tablenotetext{b}{Number of harmonics in fit including fundamental.}
\tablenotetext{c}{The 90\% confidence range is $1.6\times 10^{-6}-5.1\times 10^{-5}$ Hz.}
\tablenotetext{d}{90\% confidence upper limit.}
\end{deluxetable*}

We have investigated the presence of harmonics of the pulse frequency by including
additional sinusoids in the profile (equation \ref{profile}).  Adding first harmonic terms
decreases the Cash $\mathcal{C}$ statistic by 9.9, which has chance probability of 0.7\%.
Adding second harmonic terms decreases the Cash statistic an additional 11.4, which has a
chance probability of 0.3\%. Further harmonics continue to decrease the Cash statistic, but
these decreases are not individually as significant. The period of $5554\pm9$~s and power spectral parameters
determined with these two harmonics included are listed in Table \ref{param_tab}.

\begin{figure}[!t]
\centerline{\includegraphics[width=3.4in]{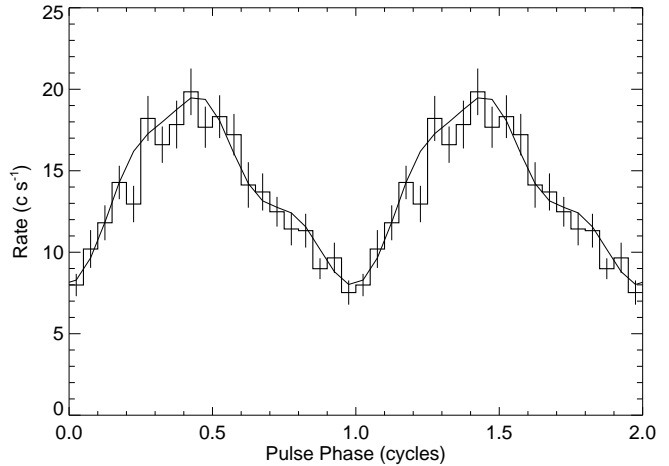}}
\caption{4U 2206+54 pulse profile obtained by epoch folding the
 Suzaku XIS 0.5-10.0 keV light curve with a period of 5554 s. The epoch of zero phase is MJD 54237.0448 (TDB).
 The smooth curve is the average of epoch folded
simulated light curves based on the maximum likelihood fit.\label{Suzaku_profile}}
\end{figure}

\begin{figure}[!b]
\centerline{\includegraphics[width=3.4in]{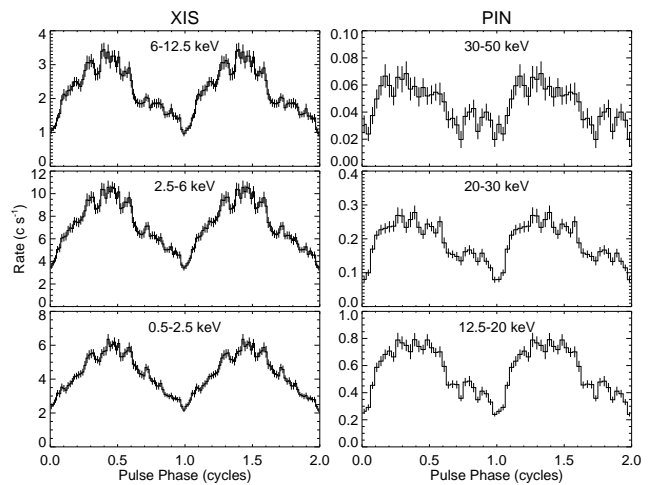}}
\caption{4U 2206+54 pulse profiles from the Suzaku XIS (on left) and HXD/PIN (on right).
The epoch of zero phase is MJD 54237.0448 (TDB).\label{Suzaku_profile2}}
\end{figure}

In Figure \ref{Suzaku_profile} we show the epoch folded Suzaku XIS rates using the
period determined from the maximum likelihood fit with two harmonics included. The errors used
are the square root of the sample variance in each phase bin divided by the number
of rates in the bin.
Using the pulse period, power spectrum, and profile parameters estimated from this same fit,
we made 10000 simulated light curves with the same binning as in the Suzaku data.
The simulation method is discussed in the appendix. The smooth curve in 
Figure \ref{Suzaku_profile} is the average epoch folded profile of these simulated light curves.

In Figure \ref{Suzaku_profile2} we show pulse profiles obtained from the Suzaku XIS and
HXD/PIN data. Background subtracted light-curves with 24\,s resolution were made for
each energy band and epoch folded with a period of 5554\,s. For the PIN profiles
light-curves where made from the cleaned event files, with live-time correction based on
144\,s averages of the pseudo events. The subtracted background was from the version 2
non-X-ray background file plus the cosmic x-ray background calculated with the flat
response per the recipe on the Suzaku Data Analysis web
page\footnote{http::/suzaku.gsfc.nasa.gov/docs/suzaku/aehp\_data\_analysis.html}. For
both XIS and PIN profiles the errors were calculated from the sample variance in each
bin. Short flares in the data has resulted in the errors being correlated between
neighboring bins.

\begin{figure}[!t]
\centerline{\includegraphics[width=3.4in]{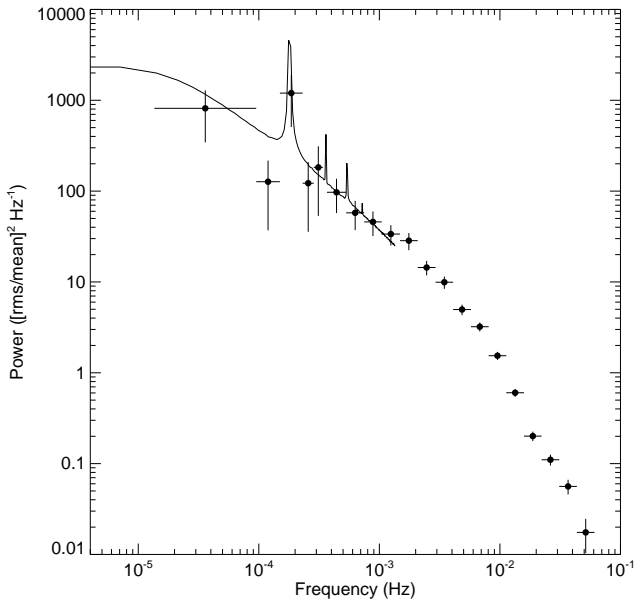}}
\caption{Power spectra estimated from a gapless 38 ks segment of the Suzaku XIS observations.
The solid curve is average power-spectra of simulated light curves  based on the maximum
likelihood fit to the Suzaku XIS light curve.\label{Suzaku_ps}}
\end{figure}

The modulation fraction of the profiles ([max-min]/[max+min]) is 49\% in 0.5-2.5 energy band, increasing to
55\% in the 20-30 keV band. A notch at the minimum is apparent in all profiles up to 30 keV. The
profile is symmetric at low energies, with the leading edge becoming stronger than the tailing edge at
higher energies.

We show in Figure \ref{Suzaku_ps} how the estimate of the pulse profile and power spectrum of
the natural log of the rates is related to the empirically determined power spectrum of the rates at
higher frequencies. For the higher frequency power spectrum we used a 37ks segment of the Suzaku XIS
observations which contained no gaps. We made a light curve
of this segment with 2~s resolution and constructed from it power spectra estimates based on the
Fourier transform of the counts.  We then produced 10000 simulated
lightcuves using the parameters estimated in the maximum likelihood fit of the Suzaku XIS data. Unlike
the previous simulations, these light curves had uniform sampling covering the duration of the
observation. The average power spectrum from these simulations is shown by the solid curve in
Figure \ref{Suzaku_ps}.

\begin{figure}[!b]
\centerline{\includegraphics[width=3.4in]{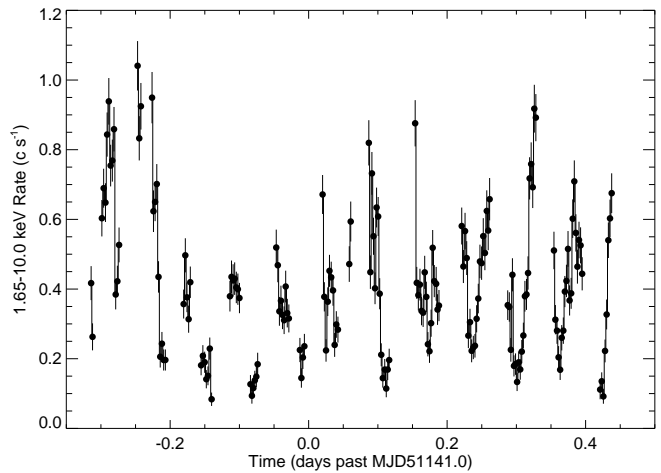}}
\caption{light-curve of 4U 2206+54 from BeppoSAX MECS in the 1.65-10 keV range.\label{SAX_lc}}
\end{figure}

\section{Analysis of the BeppoSAX light-curve}

We have reanalyzed the BeppoSAX observation presented by \citet{Masetti-etal-2004}. The source was
observed with BeppoSAX on 1998 November 11. We extracted events in the 1.65-10.0 keV range
from the version 2 merged MECS 2 and 3 files obtained from the  ASDC Multi-Mission
Interactive Archive, using a 4\arcmin\, radius extraction circle of centered on the
source. As a limit on the background, events were also extracted from an annulus with the
same area centered on a radius of 6\arcmin. The mean rate from the source region was 0.41
c s$^{-1}$ while that of the background region was 0.014 c s$^{-1}$, with much of this
likely associated with the wings of the point-spread function.

\begin{figure}[!t]
\centerline{\includegraphics[width=3.4in]{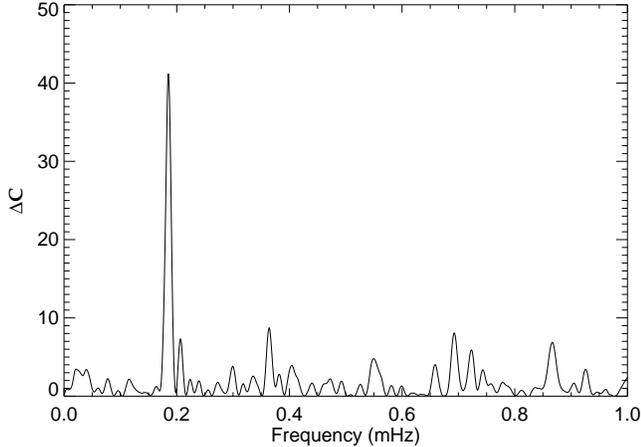}}
\caption{Delta Cash statistic versus frequency for the BeppoSAX MECS light curve.\label{SAX_deltaC}}
\end{figure}

In Figure \ref{SAX_lc} we show the
 MECS light-curve with $\sim 200$\,s binning of the 1.65-10 keV events. Again
there appears to be pulsation with a period near 0.06 days, which is most evident in the
spacing of the minima.

We show a search for long-period pulsations in Figure \ref{SAX_deltaC} using the $\Delta
\mathcal{C}$ statistic with the same modeling used for the Suzaku observation. The highest
peak has a probability of occurring by chance of $5\times 10^{-8}$ including the number
search trials in the 0-1.0 mHz interval. From the location of the highest peak we estimate
a pulse period of $5393 \pm 28$s. The peak near 0.36 mHz suggests a first harmonic. We
investigated the presence of harmonics of the pulse frequency by including additional
sinusoids in the profile (equation \ref{profile}). Adding first harmonic terms to the pulse
profile decreases the Cash statistic by 8.3 (1.6\% chance probability). Improvements from
adding additional harmonics were not as significant. A period estimate of $5420 \pm 25$~s
is obtained from the fit including the first harmonic. This and the estimated power spectra
parameters are shown in table \ref{param_tab}.

\section{Period determinations using simulated Suzaku and BeppoSAX data}

We find small but sometimes significant 
differences between the periods estimated with our maximum likelihood analysis and those
determined with the Lomb method. We also find differences between the period determined
with the maximum likelihood method when harmonics are included with the fundamental 
in the fit and the period determined when only the fundamental is included. Here we present
simulations which examine the extent to which these differences are expected. Our simulation method is presented
in the appendix.

\begin{figure}[!b]
\centerline{\includegraphics[width=3.4in]{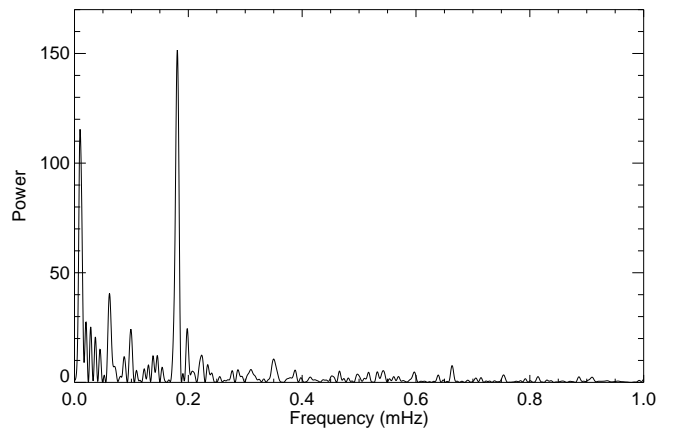}}
\caption{Pulse period search results from the Suzaku XIS data, showing the Lomb power
versus the trial pulse period.
\label{Suzaku_lomb}}
\end{figure}

In Figure \ref{Suzaku_lomb} we present the results of a search for pulsation in the
XIS light-curve using the Lomb method \citep{Lomb75}. For the Lomb power we are using twice
the normalization given in \citet{Numerical_Recipes}. A significant peak is evident near 0.18 mHz.
Using the assumptions standard for the Lomb periodigram the chance probability of this peak
to occur within the 0-1.0 mHz range due to noise is  $2\times 10^{-31}$. However these assumptions
are not correct, as we have discussed, and this significance is over-estimated. This should be compared
with Figure \ref{Suzaku_dc}. Note the decrease of the power of the lowest frequency peaks
relative to the main peak using the $\Delta C$ statistic.

The pulse period estimated from the peak of the Lomb power is 
$5541 \pm 11$\,s which is nearly consistent with our result of $5554 \pm 9$\,s using the maximum likelihood fit 
with first and second harmonic included. 
We simulated 10000 Suzaku light curves using the same times and binning as the light curve in Figure \ref{Suzaku_lc},
and the pulse period, power spectra and Fourier coefficients obtained from our maximum likelihood fit for this
light curve.  
For each we used the  Lomb method to estimate the pulse period. The period distribution was consistent with a 
Gaussian with mean displaced by $-1.8\pm2$\,s from the simulated period of 5554 s, and a standard deviation of
$17.6\pm1$\,s.  So for the Suzaku observations we conclude that the Lomb method is essentially unbiased, 
but with underestimated errors. The period measured with the maximum likelihood fit including only the fundamental
was consistent within error to our final result from the fit including first and second harmonics.

To perform the same investigation for BeppoSAX data, we made simulated light curves with the time binning of 
the MECS light curve using the estimated parameters from the maximum likelihood fit including the first harmonic. Using
10000 simulated light curves we found the distribution of the pulse period determined with the Lomb method was Gaussian 
with a mean displaced $-23.8\pm 0.3$~s  from the simulated period (5420 s) and standard deviation of 32 s. The period 
obtained using the Lomb method using the actual data is displaced by $-44$\,s from the period from our final maximum likelihood 
fit, consistent with the distribution observed in the simulations. 

The period determined with the maximum likelhood fit of the BeppoSAX data using only
the fundamental was displaced by -27\,s from the period of $5420 \pm  25$\,s determined from our final fit which 
included the first harmonic.  Using 200 of the simulated light curves, we found the distribution of
the periods determined with our maximum likelihood method with only the pulse fundamental had a mean displaced from the
simulated period (5420\,s) by $-17\pm2 $\,s and a standard deviation of $28 \pm 2$\,s.

Using 200 simulated BeepoSAX light curves we found no bias in the
periods estimated  with our maximum likelihood method with the fundamental and first harmonic
included. However the standard deviation of the estimated periods was $28.4\pm1.4$~s which larger than
the error (25\,s) we obtained from the curvature of the Cash statistic versus frequency curve for the actual data. This
is reflected in the error given in table \ref{param_tab}.

\section{Period Determinations from the EXOSAT Observations}

Observations of 4U 2206+54 where made with the Medium Energy (ME) proportional
counters on the European Space Agency's X-ray Observatory EXOSAT. 
EXOSAT/ME observed 4U 2206+54 on three occations; 1993 August 8, 1984 December 7,  and 27 June 1985. 
These observations were originally presented by \citet{Saraswat92}.

\begin{figure}[!t]
\centerline{\includegraphics[width=3.4in]{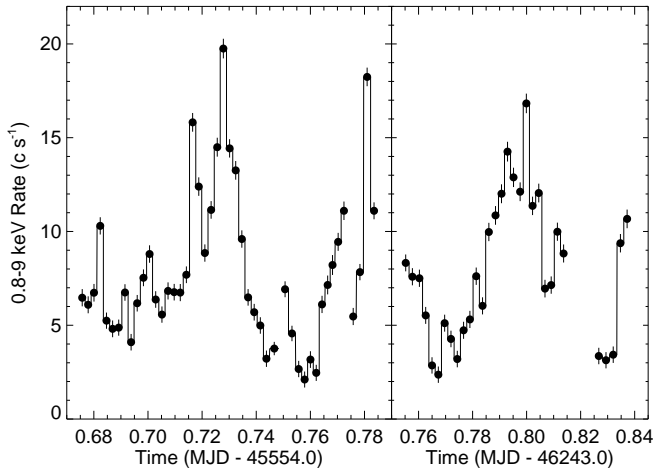}}
\caption{The EXOSAT/ME light curves for the 1983 and 1985 observations of 4U 2206+54 with 200~s binning.
\label{me_light curves}}
\end{figure}

\citet{Reig-etal-2009} analysed the ME standard product light curves from these observations, which
are available at HEASARC, and reported pulsation with a $5525 \pm 30$ period. In figure \ref{me_light curves}
we show the light curves for the first and third observation with 200~s binning. We concluded that the
background was not correctly subtracted for the standard products light curve of the second observation,
and therefore have not shown it. 

\begin{figure}[!b]
\centerline{\includegraphics[width=3.4in]{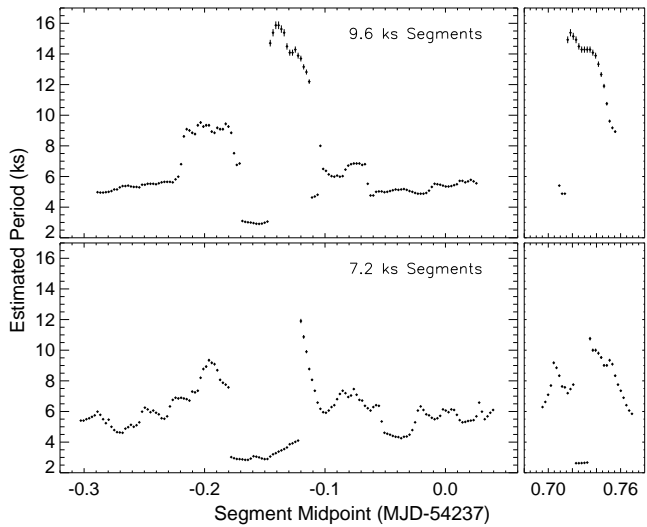}}
\caption{Periods determine using the CLEAN method using short segments of the
Suzaku XIS data. The top pannels are for 9600~ segments and the bottom pannels for 7200~s segments.
\label{clean_sim}}
\end{figure}

Given the short durations of these light curves (9630~s, 9400~s, and 7404~s respectively) relative to the pulse period, 
we do not believe a period measurement based on them is reliable. To demonstrate this we have made simulations
based on two long gapless intervals of the Suzaku XIS light curve. For the first simulation we selected a number
of overlapping 9600~s data segments within each gapless interval and from each segment made a pulse period estimate using
the Suzaku XIS rates in that segment. Since \citet{Reig-etal-2009} used the CLEAN method for period determination 
we did also. CLEAN attempts to iteratively reconstruct the Fourier amplitude spectrum of
the data \citep{Roberts-etal-1987}. We used CLEAN within the program PERIOD version 5.0-2, distributed with the 
Starlink Software Collection. In our analysis of each segment we used 5 iterations with a loop gain of 0.2.
In the top panel of figure \ref{clean_sim} we show the period from each segment versus the midpoint time of the
segment. The results of the second simulation, which used 7400~s segments, is shown in the bottom panel.
It is clear that the period estimates from these short data segments vary widely from the 5554\,s period determined
from the whole observations, and that the formal errors, which do not account for the underlying red noise, grossly
under estimate the error distribution.

\section{Discussion}

We interpret the observed X-ray pulsations of 4U~2206+54 as due to the rotation of an
accreting magnetized neutron star. An alternative interpretation, which was also
discussed for the 10000\,s pulsations of the low luminosity wind-fed system 2S
0114+650 \citep{Corbet99}, is that the observed period originates from a pulsation of
the companion BD +53$\degr$ 2790. However, this explanation seems less plausible because
such persistent modulations should be detectable in the optical photometry, but have not
been reported.
 A mechanism for transferring the modulations to the stellar wind that results
in a large modulation of the x-ray flux has not been advanced. In addition, the evolution of the
pulse profile with energy that we see in the Suzaku data would not be expected. 

Is it reasonable for us to assume that such a long period accreting
pulsar can be formed? The longest period which a neutron star can
reach in its evolution is the period at which it emerges from the
subsonic propeller state \citep{Ikhsanov-2007}
 \be\label{pbr}
p_{\rm br} = 15000\ \mu_{32}^{16/21} m^{-4/21} \left(\frac{\dot{M}_{\rm
c}}{10^{15}\,{\rm g\,s^{-1}}}\right)^{-5/7}\ {\rm s},
 \ee
where $\mu_{32}$ and $m$ are the dipole magnetic moment and mass of the neutron star in
units of $10^{32}\,{\rm G\,cm^3}$, and $1.5\,M_{\sun}$, respectively. $\dot{M}_{\rm c} =
\pi R_{\rm G}^2 \rho V_{\rm rel}$ is the mass with which a neutron star of a mass
$M_{\rm ns}$ interacts in a unit time moving through the wind of an average density
$\rho$ with a relative velocity $V_{\rm rel}$, and $R_{\rm G}=2GM_{\rm ns}/V_{\rm
rel}^2$ is the Bondi radius. For the Suzaku observations we find a $0.5-70$\,keV
luminosity of $4 \times 10^{35}\,{\rm erg\,s^{-1}}$ implying a mass accretion rate onto
the stellar surface of $\dot{M}_{\rm a} = r_{\rm ns} L_{\rm x}/GM_{\rm ns} \simeq 2
\times 10^{15}\,{\rm g\,s^{-1}}$, assuming a distance of 2.6\,kpc
\citep{Blay-etal-2006}. The mass accretion rates we inferred from published observations
all fall in the range of $4 \times 10^{13} - 3 \times 10^{15}\,{\rm g\,s^{-1}}$, with
the average near the higher end. Assuming $\dot{M}_{\rm c} \simeq \dot{M}_{\rm a}$ (i.e.
direct accretion scenario) and putting this to Eq.~\ref{pbr}, one finds that condition
$p_{\rm br} \ga 5500$\,s is satisfied only if the surface field of the neutron star in
the present epoch is $B_0 \ga 10^{14}$\,G.

An independent estimate of the field strength can be found considering the neutron star
spin evolution. The average rate of the spin frequency change between the BeppoSAX and
Suzaku observations was $\dot{\nu} = (-1.7 \pm 0.3) \times 10^{-14}\,{\rm Hz\,s^{-1}}$.
This indicates that the average spin-down torque applied to the star during this time
was $K_{\rm sd} \ga 2 \pi I |\dot{\nu}|$, where $I$ is the star's moment of inertia. If
we now adopt the canonical prescription for the spin-down torque \citep[$K_{\rm sd} =
k_{\rm t} \mu^2/r_{\rm c}^3$, see, e.g.][]{Lynden-Bell-Pringle-1974, Lipunov-1992}, one
finds $\mu \ga \mu_{\rm m}$, where
 \be\label{mum}
\mu_{\rm m} \simeq 10^{32}\ k_{\rm t}^{-1/2} m^{1/2} I_{45}^{1/2} \dot{\nu}_{-14}^{1/2}
\left(\frac{p}{5500\,{\rm s}}\right)\ {\rm G\,cm^3}.
 \ee
Here $\dot{\nu}_{-14} = |\dot{\nu}|/10^{-14}\,{\rm Hz\,s^{-1}}$, and
$I_{45}=10^{45}\,{\rm g\,cm^2}$. $r_{\rm c} = \left(GM p^2/4\pi^2\right)^{1/3}$ 
is the corotation radius of the neutron star, and $k_{\rm t}$ 
is a dimensionless parameter limited to $k_{\rm t} \la 1$. The value of
$\mu_{\rm m}$ represents the lower limit to the dipole magnetic moment of the
neutron star since the spin-up torque in the above calculations was assumed to
be negligibly small. Therefore, our estimate remains valid independently of
whether the star between the BeppoSAX and Suzaku observations was persistently
in the accretor state or its state was temporarily  changed to the propeller.
Thus, the 5500\,s pulsations in the X-ray flux of 4U~2206+54 can be
explained in terms of the spin period of the degenerate companion provided the
neutron star in this system is a magnetar whose surface field at the present
epoch exceeds $10^{14}$\,G.

The modeling of 4U~2206+54 in terms of an accretion-powered magnetar suggests
that the neutron star is relatively young (its age is limited to the
characteristic time of the supercritical magnetic field decay), with an
extended magnetosphere ($r_{\rm m} \propto \mu^{4/7}$), and a relatively small
area of hot polar caps at the base of the accretion column ($A_{\rm p} \propto
\mu^{-8/7}$). However, the assumption about spherical geometry of the accretion
flow, used for making the estimates~(\ref{pbr}) and (\ref{mum}) encounters
major difficulties in explaining the mode by which the accretion flow enters
the magnetic field of the neutron star at the magnetospheric boundary. As
first shown by \citet{Arons-Lea-1976}, and \citet{Elsner-Lamb-1977}, a steady
accretion onto the stellar surface in this case could occur only if the X-ray
luminosity of the pulsar meets the condition $L_{\rm x} > L_{\rm cr}$, where
 \be
L_{\rm cr} \simeq 10^{37}\ \mu_{32}^{1/4}\ m^{1/2}\ r_6^{-1/8}\ {\rm
erg\,s^{-1}}.
 \ee
Here $r_6$ is the radius of the neutron star in units of $10^6$\,cm.
Otherwise, the interchange instabilities of the magnetospheric boundary are
suppressed and the entry rate of accretion flow into the magnetosphere is too
small to explain the X-ray luminosity of the pulsar. The star in this case
would appear rather as a burster than a persistent source
\citep{Lamb-etal-1977}.

It is easy to see, however, that 4U~2206+54 does not meet the above criterion.
Nevertheless, it is a persistent accretion-powered X-ray pulsar with a relatively small
amplitude of variations in X-ray intensity. This contradiction may indicate that either
the spherical accretion is realized in a settling \citep{Narayan-Medvedev-2003,
Ikhsanov-2005} rather than direct accretion mode or the neutron star is accreting
material from a disk.

A scenario in which the neutron star accretes material from a hot envelope in a settling
mode on the bremsstrahlung cooling time scale implies the following condition to be
satisfied \citep{Ikhsanov-2005}
 \be
\dot{M}_{\rm c} \ga \left[\frac{t_{\rm br}(r_{\rm m})}{t_{\rm ff}(r_{\rm m})}\right]\
\frac{L_{\rm x} r_{\rm ms}}{GM_{\rm ns}}.
 \ee
Here $t_{\rm ff}(r_{\rm m})$ and $t_{\rm br}(r_{\rm m})$ are the free-fall time and
bremsstrahlung cooling time at the magnetospheric boundary. Using parameters of
4U~2206+54 one finds $M_{\rm c} \ga 10^{17}\,{\rm g\,s^{-1}}$. Therefore, the persistent
behavior of the source can be explained in terms of the settling accretion scenario only
if the strength of the wind overflowing the neutron star exceeds the typical
mass-transfer rate in a binary system with a O9.5V by almost four orders of magnitude
\citep{Reig-etal-2009}. Currently available observations do not favor the assumption
about so intensive outflow from the normal companion. However, if it were valid the
condition $p_{\rm br} > 5500$\,s could be satisfied only for the surface field of the
neutron star $\ga 4 \times 10^{15}$\,G.

It is interesting that a similar value of the magnetic field can be found considering a
situation in which the neutron star accretes material from a disk. A spin-down of the
neutron star in this case indicates that its spin period is smaller than the equilibrium
period, $p_{\rm eq}$, which is defined by equating the acceleration, $K_{\rm su} =
\dot{M}\sqrt{GM_{\rm ns}r_{\rm m}}$, and deceleration, $K_{\rm sd} = k_{\rm t}
\mu^2/r_{\rm c}^3$, torques. Using parameters of 4U~2206+54, one finds that the
condition $p \la p_{\rm eq}$ is satisfied only if the strength of the dipole field at
the stellar surface in the present epoch is limited to
 \be
B(r_{\rm ns}) \ga 3 \times 10^{15}\ \kappa_{0.5}^{7/24} k_{\rm t}^{-7/12} m^{5/6}
\dot{M}_{15.3}^{1/2} \left(\frac{p}{5500\,{\rm s}}\right)^{7/6}\ {\rm G},
 \ee
where $\kappa_{0.5}=\kappa/0.5$ is the parameter accounting the geometry of the
accretion flow normalized following \citet{Ghosh-Lamb-1978}. The spin-down time scale of
a newly formed neutron star to a period of 5500\,s under these conditions is close to
5000\,yr \citep[see Eq.~20 in][]{Ikhsanov-2007}. This exceeds the decay time of the
magnetic field confined to the crust of the neutron star \citep{Colpi-etal-2000}, but is
still smaller than the decay time of the field permeating the core. Since the X-ray
spectrum of the source resembles spectra of accretion-powered pulsars one can assume
that the contribution of the source associated with the field decay to the system X-ray
luminosity is small. The field decay time in this case can be limited to
 \be
t_{\rm dec} \ga 2 \times 10^5\ \mu_{33.3}^2\ r_6^{-3}\ L_{35.6}^{-1}~{\rm yr},
 \ee
which is consistent within results of \citet{Heyl-Kulkarni-1998}, who treated the
magneto-thermal evolution of the star in terms of ambipolar diffusion. Here $L_{35.6}$
is the X-ray luminosity of the source express in unites of $4 \times 10^{35}\,{\rm
erg\,s^{-1}}$. The proton cyclotron feature in this case is expected to be observed at
the energy of $\sim 30\ (B_0/5 \times 10^{15}\,{\rm G})$\,keV.

It, therefore, appears that our interpretation of the observed 5500\,s pulsations in
terms of spin period of the neutron star is reasonable and suggests that the neutron
star in 4U~2206+54 is a magnetar with a surface field of about $(3-5)\times 10^{15}$\,G.
A question about the geometry of the accretion flow remains, however, open so far. The
spherical approximation of the flow geometry can be applied only under the condition
$\dot{M}_{\rm c} \gg \dot{M}_{\rm a}$. On the other hand, the observed wind velocity of
the O-star companion, $V_{\rm w} \ga 350\,{\rm km\,s^{-1}}$,
\citep[see][]{Ribo-etal-2006}, substantially exceeds the upper limit to the relative
velocity between the star and the wind at which the disk formation could be expected,
  \be\label{vrel}
V_{\rm rel} \la V_{\rm cr} \simeq 120\ \xi_{0.2}^{1/4}\ \mu_{33}^{-1/14}\
m^{11/28}\ \dot{M}_{15}^{1/28}\ P_{20}^{-1/4}\ {\rm km\,s^{-1}}.
 \ee
Here $P_{20}$ is the orbital period of the system in units of $20$\,days, and
$\xi_{0.2}=\xi/0.2$ is the parameter accounting for the inhomogeneities of the accretion
flow, which is normalized following \citet{Ruffert-1999}. Nevertheless, one cannot
exclude a presence of a fossil disk surrounding the magnetosphere of the neutron star
which could be formed at the fall-back stage of the magnetar evolutionary track
\citep{Woosley-1988, Eksi-Alpar-2003}. Further studies of appearance of such a disk in a
situation when the neutron star rotates extremely slowly and has a huge magnetosphere
could help to choose appropriate flow geometry and reconstruct the accretion picture in
this enigmatic source.

Finally, we would like to point out here that at the rate $\dot{\nu} = (-1.7 \pm 0.3)
\times 10^{-14}\,{\rm Hz\,s^{-1}}$ the pulsar would reach zero frequency in 300\,yr,
making it is unlikely that this is a long-term trend. Over long time-scales wind-fed
pulsars tend to show random-walk behavior in their spin frequency. For 4U 2206+54 we can
estimate the random walk strength as $S=~<(\Delta \nu)^2/\Delta t>~\simeq~8 \times
10^{-20}\,{\rm Hz^2\,s^{-1}}$. For comparison Vela X-1 has a random walk strength of $2
\times 10^{-20}$ \citep{Deeter87}. The random walk behavior must break down, and the
torque become correlated at short time-scales since the spin-up rate $\dot{\nu} \sim
(S/\Delta t)^{1/2}$ from wind accretion is limited by that due to disk accretion at the
same mass accretion rate. Assuming a mean mass accretion rate of $10^{15}\,{\rm
g\,s^{-1}}$, and a magnetic field of $10^{15}$\,G, we find the spin-up rate must be
correlated on timescales shorter than $\sim 5$\,days.

Based on the strong emission in Helium lines \citet{Negueruela01} and
\citet{Blay-etal-2006} have proposed that the optical companion BD +53$\degr$ 2790 is
similar to $\theta^1$\,Ori\,C, which has a oblique magnetic dipole with kG surface
strength which magnetically channels its stellar wind. If so, this long correlation
timescale could be related to long-term correlations in the wind speed or density
associated with this magnetic channeling.

\acknowledgments

M.H.F. acknowledges support from NASA grant NNX08AG12G. N.R.I acknowledges supported
from NASA Postdoctoral Program at NASA Marshall Space Flight Center, administered by Oak
Ridge Associated Universities through a contract with NASA, and support from Russian
Foundation of Basic Research under the grant 07-02-00535a.

\appendix
\section{Calculation of the Cash statistics $\mathcal{C}_0$ and $\mathcal{C}_1$}
Numerically we calculate $V$ in equation \ref{covar} as
\begin{equation}
V_{kl} = \sigma_k^2\delta_{kl}
       +\sum_{j=0}^M w_j \mathcal{S}(f_j) {\rm sinc}^2(\pi f_j \tau)
      \cos( 2\pi f_j [t_k-t_l]) \Delta f ~~.
\label{Vsum}\end{equation}
Here the frequency $f_j = j\Delta f$ and the frequency step used is $\Delta f$ = $(16T)^{-1}$
with $T$ the duration of the observation. The weigths $w_j$ give trapazoidal integration with $w_j = 1$
except for $w_0 = w_N = 0.5$. $M$ is chosen as the nearest integer to $T/\tau$ so that the integral
extends to where the power is cut off by the binning effect.

By using
\begin{equation}
\cos( 2\pi f_j [t_k-t_l]) = \cos  2\pi f_j t_k \cos  2\pi f_j t_l
                           +\sin  2\pi f_j t_k \sin  2\pi f_j t_l
\end{equation}
equation \ref{Vsum} can be express as
\begin{equation}
V = U^T U \label{Vfactor}
\end{equation}
where the matrix $U$ has $N+2M+1$ rows and $N$ columns, with $N$ is the number of measurements, and $T$ signifies the matrix
transpose. Explicitly $U$ is given by
\begin{equation}
  \begin{array}{l} U_{i,k}\vphantom{]^\frac{1}{2}} \\ U_{N,k}\vphantom{]^\frac{1}{2}} \\
           U_{N+2j-1,k}\vphantom{]^\frac{1}{2}} \\
           U_{N+2j,k}\vphantom{]^\frac{1}{2}} \end{array}
  \begin{array}{c} =\vphantom{]^\frac{1}{2}} \\ =\vphantom{]^\frac{1}{2}} \\
                   =\vphantom{]^\frac{1}{2}} \\ =\vphantom{]^\frac{1}{2}}
  \end{array}
  \begin{array}{l}
     \sigma_i \delta_{i,k}\vphantom{]^\frac{1}{2}} \\
      {[w_0 \mathcal{S}(0) \Delta f]^\frac{1}{2}}                 \\
      {[w_j \mathcal{S}(f_j) {\rm sinc}^2(\pi f_j \tau)\Delta f]}^\frac{1}{2}\cos  2\pi f_j t_k \\
      {[w_j \mathcal{S}(f_j) {\rm sinc}^2(\pi f_j \tau)\Delta f]}^\frac{1}{2}\sin  2\pi f_j t_k
  \end{array}
   \begin{array}{l}
      {\rm for~~} 0 \leq i < N\vphantom{]^\frac{1}{2}} \\
      \vphantom{]^\frac{1}{2}}\\
       {\rm for~~} 1 \leq j \le M \vphantom{]^\frac{1}{2}}\\
       {\rm for~~} 1 \leq j \le M \vphantom{]^\frac{1}{2}}
   \end{array}
\end{equation}
By applying a series of N orthgonal Householder transformations (reflections) $U$ can be transformed into
an $N \times N$ upper triangular matrix $\tilde{U}$, while perserving equation \ref{Vfactor} \citep{Householder58}.
We then have
\begin{equation}
 \mathcal{C}_1 = |\tilde{U}^{-T}(\mathbf{y}-\mathbf{p})|^2~+2\sum_i {\rm ln}(|\tilde{U}_{ii}|)~, \label{C1a}
\end{equation}
with a simular expression for $\mathcal{C}_0$. Here $\mathbf{y}$ is the vector with components $y_k$ and $\mathbf{p}$ is the
vector with components $p_k$. We will introduce the augmented parameter vector ${\bf q} = [\mu,a,b,-1]$, and using
equation \ref{profile} define the matrix $H$ so that $\mathbf{p}-\mathbf{y} = H{\bf q}$. Then with $R=\tilde{U}^{-T}H$ we
have
\begin{equation}
 \mathcal{C}_1 = |R{\bf q}|^2~+2\sum_i {\rm ln}(|\tilde{U}_{ii}|)~, \label{C1b}
\end{equation}
By applying Householder transformations $R$ can be transformed into a $4\times 4$ upper
triangular matrix $\tilde{R}$ while preserving equation \ref{C1b}. We can then minimize $\mathcal{C}_1$
with respect to $\mu$, $a$ and $b$ by setting the first three components of $\tilde{R}{\bf q}$ to
zero and solving for $\mu$, $a$ and $b$, with the minimum of $|\tilde{R}{\bf q}|^2$ given by $R_{33}^2$.

We minimize $\mathcal{C}_0$ and $\mathcal{C}_1$ with respect to the power spectral parameters $s_0$, $a_0$, and $\Gamma$
by the downhill simplex method.

\section{Light Curve Simulations}

In our simulations we first simulate natural log rates $y_k$ with covariance given by equation \ref{Vsum} with the
measurement errors $\sigma_k$ set to zero, and means $p_k$ given by equation \ref{profile} (perhaps with 
terms for harmonics).  Simulated counting statistical errors are then added to the rates $r_k$. 

As above an upper triangular matrix $\tilde{U}$ is calculated such that $\tilde{U}^T\tilde{U} = V$. 
Then with  $\mathbf{x}$ is vector of $N$ random normal varients with unit variance zero mean we calculate
\begin{equation}
  \mathbf{\Psi} = \tilde{U}^T\mathbf{x}
\end{equation}
The covariance matrix of $\mathbf{\Psi}$ is
\begin{equation}
   <\mathbf{\Psi}\mathbf{\Psi}^T> =  \tilde{U}^T<\mathbf{x}\mathbf{x}^T>\tilde{U} =  \tilde{U}^T\tilde{U} = V~~.
\end{equation}
The simulated rates prior (without counting statistical errors) are given by 
\begin{equation}
{r_k}  = \exp({\Psi_k}+{p_k})~~.
\end{equation}
Counting statistical errors are then simulated from these rates and the bin widths.


\begin{thebibliography}{}

\bibitem[Arons \& Lea(1976)]{Arons-Lea-1976}
 Arons, J., Lea, S.M. 1976, ApJ, 207, 914

\bibitem[Blay et al.(2005)]{Blay05a} Blay, P., Rib\'{o}, M., Negueruela, I.,
Torrej\'{o}n, J. M., Reig, P., Camero, A., Mirabel, I. F. \& Reglero, V.
2005, \aap, 438, 963

\bibitem[Blay(2005)]{Blay05b} Blay, Pere 2005, Ph. D. Thesis, Universitat de Val\`{e}ncia, Spain

\bibitem[Blay et~al.(2006)]{Blay-etal-2006}
Blay, P. et~al. 2006, \aap, 446, 1095


\bibitem[Cash (1979)]{Cash79} Cash, W. 1979, \apj, 228, 939

\bibitem[Colpi et~al.(2000)]{Colpi-etal-2000}
 Colpi, M., Geppert, U., Page, D. 2000, ApJ, 529, L29


\bibitem[Corbet et al.(1999)]{Corbet99} Corbet, R. H. D., Finley, J. P., \& Peel, A. G. 1999,
\apj 511, 876

\bibitem[Corbet \& Peele(2001)]{Corbet01} Corbet, R. H. D. \& Peele, A. G. 2001,
\apj, 562, 936

\bibitem[Corbet et al.(2007)]{Corbet07} Corbet, R. H. D., Markwardt, C. B., \& Tueller, J. 2007,
\apj 655, 458

\bibitem[Deeter et al.(1987)]{Deeter87} Deeter, J. E., Boynton, P. E., Shibazaki, N. et al. 1987,
Astron. J. 93, 877


\bibitem[Elsner \& Lamb(1977)]{Elsner-Lamb-1977}
 Elsner, R.F., \& Lamb, F.K. 1977, \apj, 215, 897

\bibitem[Ek\c{s}i \& Alpar(2003)]{Eksi-Alpar-2003}
  Ek\c{s}i, K.J., \& Alpar, M.A. 2003, \apj, 599, 450

 \bibitem[Ghosh \& Lamb(1978)]{Ghosh-Lamb-1978}
 Ghosh, P., Lamb, F.K. 1978, ApJ, 223, L83


\bibitem[Householder(1958)]{Householder58}Householder, A. S. 1958, J. Assoc. Comp. Mach. 5, 339

\bibitem[Heyl \& Kulkarni(1998)]{Heyl-Kulkarni-1998}
  Heyl, J.S., \& Kulkarni, S.R. 1998, \apj, 506, L61

\bibitem[Ikhsanov(2005)]{Ikhsanov-2005}
 Ikhsanov, N.R. 2005, Astron. Lett., 31, 586

\bibitem[Ikhsanov(2007)]{Ikhsanov-2007}
  Ikhsanov, N.R. 2007, \mnras, 375, 698

\bibitem[Komaya et al.(2007)]{Komaya07}Komaya, K. et al. 2007, PASJ, 59, 23

\bibitem[Lamb et~al.(1977)]{Lamb-etal-1977}
 Lamb, F.K., Fabian, A.C., Pringle, J.E., \& Lamb, D.Q. 1977, \apj, 217, 197

\bibitem[Lynden-Bell \& Pringle(1974)]{Lynden-Bell-Pringle-1974}
 Lynden-Bell, D., Pringle, J.E. 1974, MNRAS, 168, 603

\bibitem[Lipunov(1992)]{Lipunov-1992}
 Lipunov, V.M. 1992, Astrophysics of neutron stars,
 Springer-Verlag, Heidelberg

\bibitem[Lomb(1975)]{Lomb75}Lomb, N. R. 1975, \apss, 39, 447

\bibitem[Masetti et~al.(2004)]{Masetti-etal-2004}
 Masetti, N., et~al. 2004, \aap, 423, 311

\bibitem[Narayan \& Medvedev(2003)]{Narayan-Medvedev-2003}
 Narayan, R. \& Medvedev, M.V. 2003, \mnras, 343, 1007


\bibitem[Negueruela \& Reig(2001)]{Negueruela01} Negueruela, I. \& Reig, P. 2001,
\aap, 371, 1056

\bibitem[Press et al. (1992)]{Numerical_Recipes} Press, W. H. et al., Numerical Recipes in Fortran, Second Ed.,
1992, Cambridge U. Press, p. 569

\bibitem[Reig et~al.(2009)]{Reig-etal-2009}
 Reig, P., et~al. 2009, \aap, 494, 1073

\bibitem[Rib\'o et~al.(2006)]{Ribo-etal-2006}
 Rib\'o, M., et~al. 2006, \aap, 449, 687

\bibitem[Roberts et~al.(1987)]{Roberts-etal-1987}
Roberts, D.~H., Leh\'{a}r, J. \& Dreher, J.~W. 1987, \aj, 93, 968

\bibitem[Ruffert(1999)]{Ruffert-1999}
 Ruffert, M. 1999, \aap, 346, 861

\bibitem[Saraswat \& Apparao(1992)]{Saraswat92} Saraswat, P. \& Apparao, K. M. V. 1992,
\apj, 401, 678

\bibitem[Takahashi et al.(2007)]{Takahashi07}Takahashi, T. et al. PASJ, 59, 35

\bibitem[Torrej\'on et~al.(2004)]{Torrejon-etal-2004}
 Torrej\'on, J.M., et~al. 2004, \aap, 423, 301

\bibitem[Woosley(1988)]{Woosley-1988}
 Woosley, S.E. 1988, \apj, 330, 218

\end{thebibliography}
\end{document}